# Strong evidence for *d*-electron spin transport at room temperature at a LaAlO$_3$/SrTiO$_3$ interface

Ryo Ohshima [1,2], Yuichiro Ando [1], Kosuke Matsuzaki [3], Tomofumi Susaki [3], Mathias Weiler [4,5], Stefan Klingler [4,5], Hans Huebl [4,5,6], Eiji Shikoh [7], Teruya Shinjo [1], Sebastian T. B. Goennenwein [4,5,6] and Masashi Shiraishi [1,*]

1. Department of Electronic Science and Engineering, Kyoto Univ., 615-8510 Kyoto, Japan.
2. Graduate School of Engineering Science, Osaka Univ., 560-8530 Toyonaka, Japan.
3. Secure Materials Center, Materials and Structures Lab., Tokyo Institute of Technology, 226-8503 Yokohama, Japan.
4. Walther-Meißner-Institut, Bayerische Akademie der Wissenschaften, 85748 Garching, Germany.
5. Physik-Department, Technische Universität München, 85748 Garching, Germany
6. Nanosystems Initiative Munich (NIM), Schellingstraße 4, 80799 München, Germany
7. Graduate School of Engineering, Osaka City Univ., 558-8585 Osaka, Japan.

A *d*-orbital electron has an anisotropic electron orbital and is a source of magnetism. The realization of a 2-dimensional electron gas (2DEG) embedded at a $LaAlO_3/SrTiO_3$ interface surprised researchers in materials and physical sciences because the 2DEG consists of 3*d*-electrons of Ti with extraordinarily large carrier mobility, even in the insulating oxide heterostructure. To date, a wide variety of physical phenomena, such as ferromagnetism and the quantum Hall effect, have been discovered in this 2DEG systems, demonstrating the ability of the *d*-electron 2DEG systems to provide a material platform for the study of interesting physics. However, because of both ferromagnetism and the Rashba field, long-range spin transport and the exploitation of spintronics functions have been believed difficult to implement in the *d*-electron 2DEG systems. Here, we report the experimental demonstration of room-temperature spin transport in the *d*-electron-based 2DEG at a $LaAlO_3/SrTiO_3$ interface, where the spin relaxation length is ca. 300 nm. Our finding, which counters the conventional understandings to *d*-electron 2DEGs, opens a new field of *d*-electron spintronics. Furthermore, this work highlights a novel spin function in the conductive oxide system.

A two-dimensional electron gas (2DEG) generated between LaAlO$_3$ (LAO) and SrTiO$_3$ (STO) consists of itinerant $d$-orbital electrons in the $t_{2g}$ orbitals of Ti in STO[1,2]. This anisotropic orbit induces ferromagnetism, superconductivity, and even their coexistence at LAO/STO interfaces[3–8]. The subbands of LAO/STO interfaces have a complicated structure and cross each other due to the strong orbital hybridization[9–12], resulting in the enhancement of spin-orbit coupling and nonmonotonic gate tunability[13–16]. It is believed that long-rage spin transport in the 2DEGs at LAO/STO interfaces is difficult to achieve due to the Rashba spin-orbit interactions and their $d$-electrons character. In fact, the spin diffusion length of 3$d$-ferromagnetic metals is typically less than 10 nm at room temperature (RT)[17,18]. However, enabling spintronics functions in a $d$-electron 2DEG is very attractive because theoretical calculations have predicted a long spin life time of ca. 500 ns for a LAO/STO interface, even at room temperature[19]. Thus, an experimental demonstration of lateral spin transport in the 2DEG formed by $d$-orbital electrons at the LAO/STO interface is a key challenge and would represent a significant advance.

To date, there is no report of spin transport in a 2DEG at LAO/STO interfaces. The first major step towards spintronics, namely to inject and accumulate spins into a $d$-electron 2DEG, was reported in 2012, through an electrical three-terminal method[20]. However, after a long-term, heated discussion on the validity of the method[21], it has recently been concluded that signals measured with this method cannot be attributed to spin accumulation and are due instead to spurious effects, such as spin traps in tunnel barriers[22,23]. Thus, a spintronics study in a $d$-electron 2DEG system remains to be accomplished. In this letter, we demonstrate the first successful lateral spin transport in the 2DEG formed at the LAO/STO interface at RT, i.e. the first realization of $d$-electron spin transport. In our experiments, we inject a spin current into the 2DEG by means of spin-pumping from a NiFe (Py) stripe and use a non-local inverse spin Hall effect detection using a Pt or Ta stripe. The spin diffusion length was estimated to be ca. 300 nm by a conventional simple diffusion model that takes into account the geometrical effects. Our study opens a new frontier of semiconductor spintronics, where *s-p* hybridized electrons mainly carry a spin degree of freedom.

A schematic of a typical sample structure is shown in Fig. 1a. A LAO film with a thickness of 5 unit cells (u.c.), was grown on a STO substrate by pulsed laser deposition. For this LAO thickness, a 2DEG indeed forms at the interface[24]. Via electron beam lithography, electron beam deposition, sputtering and lift-off, we defined two parallel strips made from Py and Pt (or Ta), respectively, on the LAO layer and they are separated by a distance *L*. The sample with the Ta electrode was prepared for control experiments by changing the polarity of the spin Hall angle. The details of the sample preparation procedure are described in the *Methods* section. A bias voltage between the Py (or the Ta) and the detector electrode was applied to measure a current-voltage (*I-V*) characteristic (see Figs. 1b and 1c). It is well known that STO substrate is doped by using a milling process[25]. The Pt and Ta electrodes in this study were prepared by using a high-power sputtering process, resulting in the pseudo-Ohmic contact at the Py-LAO/STO interface. The pseudo-Ohmic contact allows injecting spins from the Py and also detecting spin transport signals in the detector electrode. We applied a microwave ($f$ = 9.63 GHz) and a static magnetic field with angle $\theta_H$ to the +*x*-axis (see Fig. 1a) to the Py strip on the LAO layer in order to excite the ferromagnetic resonance (FMR) of the Py. Under condition of the FMR, spin angular momentum is transferred from the Py to the 2DEG, resulting in the generation of pure spin current in the 2DEG[26,27]. The pure spin current in the 2DEG was detected by using the independent, separate NM (=Pt or Ta) strip deposited on the LAO layer. The propagating pure spin current is absorbed in the NM and converted to a charge current by the inverse spin Hall effect (ISHE) of the NM[28], generating an electromotive force (EMF) in it (see also the *Methods* section). In this experiment, spin current flowing into the Pt along +*y*-axis generates the negative EMF in the Pt at $\theta_H = 0°$. Figures 2a and 2d show a simultaneously recorded FMR spectrum of the Py and an EMF from the Pt, respectively, when *L* was set to be 890 nm. The EMF appears under the FMR. The EMF spectrum consists of symmetric and asymmetric components. The symmetric part $V_{sym}$ is attributed to the ISHE and the asymmetric part $V_{asym}$ is due to the anomalous Hall effect of the Py[28] (see Supplemental Information for details); the symmetric component is described in Fig. 2d.

The ISHE has a characteristic symmetry in a relationship between the pure spin current direction $\boldsymbol{J_S}$, the charge current direction $\boldsymbol{J_C}$ and the spin polarization vector $\boldsymbol{\sigma}$, i.e.

$$\boldsymbol{J_C} = \theta_{SHE}\left(\frac{2e}{\hbar}\right)\boldsymbol{J_S} \times \boldsymbol{\sigma}, \tag{1}$$

where $\theta_{SHE}$ is the spin Hall angle, $e$ is the electronic charge and $\hbar$ is the Dirac constant[28]. The symmetry in equation (1) indicates that the polarity of the EMF is reversed when the direction of the external magnetic field, i.e., the direction of the spin polarization vector of the pure spin current is reversed. Figs. 2a-c and 2d-e show the FMR spectrum and EMF at $\theta_H = 0°$, 90° and 180°, respectively. The $\theta_H$ dependence of the EMFs is consistent with the symmetry of the ISHE shown in equation (1): the polarities of electromotive forces at $\theta_H = 0°$ and 180° were reversed, and the signal at $\theta_H = 90°$ disappeared. These results strongly suggest that the observed EMFs under the FMR conditions can be ascribed to the ISHE in the Pt, i.e., the propagation of pure spin current in the 2DEG at RT. Deconvolutions of the EMFs were performed for $\theta_H = 0°$ and 180° and we used $(V_{sym}(\theta_H = 0) - V_{sym}(\theta_H = 180))/2 = -1.16$ μV, for the component of $V_{ISHE}$ that changes sign under H-inversion. The non-negligible asymmetric contribution is attributable to spurious effects, such as the anomalous Hall effect of the Py[28] and the metal-capping-induced ferromagnetism of the 2DEG[29].

However, utmost care is necessary to conclude that successful spin transport has occurred, because several spurious effects generating EMFs from conductive ferromagnets under FMR have been reported[30–33]. We implemented the following control experiment to confirm that the measured EMFs indeed evidence of a successful, long-range propagation of the pure spin currents in the 2DEG. The control experiment was to replace the detector electrode material. If the EMF would arise via FMR rectification or similar effects within the ferromagnet itself, then the polarity of the EMF would be independent of the ISHE detector materials. We therefore changed the ISHE detector material from Pt to Ta with the different sign of the spin Hall angle[34–36]. Accordingly, an EMF arising due to the propagation of a pure spin current in the 2DEG and ISHE detection in the non-magnetic metal strip should invert sign upon replacing Pt with Ta, while no EMF sign change would occur if the EMF would stem from the ferromagnetic strip. The results on the introduction of the Ta detector are shown in Figs. 3d-f, and a comparison between the result on Ta and Pt is shown in Table I. Notably, the sign of $V_{ISHE}$ obtained using a Ta electrode was opposite to that of Pt, which strongly supports the notion that long-range spin transport in the 2DEG indeed has been successfully achieved. It is important to note that the Py and Pt or Ta strips are only connected via the 2DEG at the LAO/STO interface, since the STO and the LAO layers are insulators. The amplitudes of the symmetry component of the electromotive forces in the cases of $\theta_H = 0°$ and $180°$ were different, and we think the reason is the heating of the Py under the FMR. In our current experimental set-up, we tried to set the sample in the center of the cavity and excited the FMR via microwave with fixed frequency and swept a static (dc) magnetic field. However, the position of the sample was not completely the center of the Py, and the Py had a thermal gradient along the detector strip (Pt and Ta), which can generate an electromotive force under the FMR. Here, it should be noted that this is not a spin-dependent effect and also that the spin-dependent thermal signals cannot be superimposed because the direction of the spin-dependent thermal signal should be perpendicular to the Py film plane judging from the directions of the thermal gradient and the external magnetic field. It is known that a spin-independent electromotive force does not change its sign when the angle of the static magnetic field is changed from 0° to 180° as reported by Ando et al[37]. In our experiment, the Lorentzian-type signals due to this spin-independent thermal effect were superimposed to the spin-transport-induced electromotive forces as shown in Figs. 2d, 2f and Figs. 3d, 3f, resulting in the difference of the amplitude of the symmetry component between 0° to 180° (in fact, we estimated the thermal contribution for the sample with Pt and that with the Ta to be -0.6 and -1.7 µV, respectively, and the non-equality problem of the magnitude of the electromotive forces is solved by subtracting the contributions). At 90° (in Figs. 2e and 3e), the absorption of the microwave, $I$, was smaller than that at 0° to 180° ($I(\theta_H = 90) = 2400$ and $I(\theta_H = 0) = 9000$, estimated from the integration of FMR spectra) and it is concluded that the thermal effect was small enough. In the following paragraphs, we will discuss spin diffusion length of the 2DEG, where the spin-independent thermal signals were eliminated from the original spectra, i.e., we discuss the symmetric Lorentzian component that change their polarity by changing the electrode type from Pt to Ta.

The combination of both experimental results shows that spin transport in the LAO/STO 2DEG indeed is possible, over surprisingly long distances of several 100 nm and allowed us to estimate the spin diffusion length of the LAO/STO interfaces $\lambda_{2DEG}$. We used a well-established spin diffusion model including the contribution of sample geometry for the estimation[38–40]. For the Pt (Ta) electrode sample, the injected spin current density $j_S^{Py}$ is estimated to be $7.44 \times 10^{-10}$ Jm$^{-2}$ ($8.43 \times 10^{-10}$ Jm$^{-2}$) by calculating the real part of the mixing conductance ($g_r^{\uparrow\downarrow} = 1.78 \times 10^{-19}$ m$^{-2}$ ($1.97 \times 10^{-19}$ m$^{-2}$)), see also the Supplemental Information). The absorbed spin current density $j_S^{NM}$ (NM = Pt or Ta) can be calculated by integrating from the center to the edge of the Py wire[39]. The sheet conductance of the LAO/STO interfaces $\sigma_{2DEG}$ was measured to be $2.58\times10^{-5}$ 1/Ω by using the van der Pauw method; this value is much smaller than the sheet conductance of Pt ($1.54\times10^{-2}$ 1/Ω) and Ta ($2.72\times10^{-3}$ 1/Ω). Next, we constructed the equation for the EMF obtained from Pt (Ta) electrode as[39]

$$V_{\text{ISHE}} = \frac{l_{\text{Py}}|\theta_{\text{SHE}}^{\text{NM}}|\lambda_{\text{NM}}\tanh(d_{\text{NM}}/2\lambda_{\text{NM}})}{d_{\text{NM}}\sigma_{\text{NM}}}\left(\frac{2e}{\hbar}\right)j_S^{\text{NM}},$$
$$j_S^{\text{NM}} = j_S^{\text{Py}} e^{-\left(\frac{L}{\lambda_{2DEG}}\right)}\frac{\lambda_{2DEG}}{w_{\text{NM}}}\left(1 - e^{-\left(\frac{w_{\text{Py}}}{2\lambda_{2DEG}}\right)}\right),$$
(2)

where $l_{\text{Py}} = 900$ μm is the length of the Py wire, $|\theta_{\text{SHE}}^{\text{Pt}}| = 1.0\times10^{-1}$ ($|\theta_{\text{SHE}}^{\text{Ta}}| = 7.1\times10^{-2}$) is the spin Hall angle of Pt (Ta)[36], $\lambda_{\text{Pt}} = 7.3$ nm ($\lambda_{\text{Ta}} = 1.9$ nm) is the spin diffusion length of Pt (Ta)[36], $d_{\text{Ta}} = 7.4$ nm ($d_{\text{Pt}} = 7.9$ nm) is the thickness of Pt (Ta), $\sigma_{\text{Pt}} = 2.1\times10^6$ 1/Ωm ($\sigma_{\text{Ta}} = 3.5\times10^5$ 1/Ωm) is the conductivity of Pt (Ta)[36], $w_{\text{NM}} = 1$ μm and $w_{\text{Py}} = 200$ μm are the NM and Py electrode widths, respectively, and $L = 890$ nm (480 nm) is the gap length between the Py and Pt (Ta) electrodes. $V_{\text{ISHE}} = 1.16$ μV (1.74 μV) is the EMF of the ISHE obtained from the Pt (Ta) electrode. Consequently, the spin diffusion length of the 2DEG at RT is estimated to be 340 nm and 210 nm in the case of the Pt and the Ta detectors. Spin diffusion length was also estimated by changing the gap length between spin injection and detection electrodes as crosschecking of the estimation. Spin transport signal is exponentially decreased with increasing the transport distance by the factor of spin diffusion length, $\lambda_{2DEG}$, in the diffusion regime. We measured the ISHE-induced electric current, $I_{\text{ISHE}} = V_{\text{ISHE}}/R$, by changing the distance from the Py to the detector electrodes (Pt and Ta), $L$. To normalize the ISHE-induced electric current in the different detector materials (Pt and Ta), $I_{\text{norm}} = I_{\text{ISHE}}/|\theta_{\text{SHE}}^{\text{NM}}|\lambda_{\text{NM}}\tanh(d/2\lambda_{\text{NM}})$ was used in the transport distance dependence because the spin diffusion lengths and spin Hall angles $\theta_{\text{SHE}}$ of Pt and Ta are different. We fit the normalized ISHE-induced electric current by using an equation, $I_{\text{norm}} = a\exp(-L/\lambda_{2DEG})$,

where $a$ is a constant. The symmetric part of the electromotive forces exhibits exponential dependence as a function of the gap length between the Py and the Pt (or the Ta). The spin diffusion length of the LAO/STO interface is estimated to be $337 \pm 181$ nm from the gap length dependence (see Fig. 4). The difference of the calculated values of the spin diffusion length (340 nm for the Pt equipped sample and 210 nm for the Ta equipped sample) is within the range of the experimental error bar, and the difference is ascribable to experimental uncertainty. Furthermore, we estimate $\lambda_{2DEG}$ to be 360 nm from another model in the case of the Pt detector (see also the Supplementary Information). The good agreement of the results also supports the demonstration of the spin transport and the appropriate use of the model to describe the spin transport. The following findings are notable: (1) the estimated values $\lambda_{2DEG}$ were equivalent, and (2) the sign reversal of the EMF for changing Pt to Ta was observed. The diffusion constant of LAO/STO interfaces $D$ was estimated to be 0.18 cm$^2$/s (see the Supplemental Information), and the spin lifetime $\tau_S$ was estimated to be 6.4 ns and 2.5 ns from the relation, $\lambda_{2DEG} = \sqrt{D\tau_S}$, in the case of the Pt and the Ta detectors, respectively.

The long spin diffusion length of itinerant $d$-electrons can be confirmed by the measurement of the Gilbert damping by studying the frequency dependence of the Py FMR line width in the Py/LAO/STO system[41,42]. In metal bilayer systems, the Gilbert damping parameter $\alpha$ is enhanced compared with that from a single film. We fit the data recorded for the full width at half maximum $\Delta H$ to $\mu_0 \Delta H = \mu_0 \Delta H_0 + 4\pi\alpha f/\gamma$, where $\Delta H_0$ is the factor of long-range magnetic inhomogeneities, $f$ is the microwave frequency and $\gamma$ is the gyromagnetic ratio[41,42]. This method allows separating inhomogenous broadening and damping contributions to $\Delta H$. Note that the enhancement of $\alpha$ is suppressed when the spin sink of this bilayer is good spin translator such as Cu because there is little spin dissipation[43]. The Py film was evaporated on the LAO/STO samples with 2, 5, and 10 u.c. of the LAO film, in addition to the Py film on STO substrate, and the frequency dependence was measured (the details are presented in the *Methods* section). Here, the Py fully covered the LAO/STO and the STO surface, which is a dominant difference from the structure of the spin transport device (the Py spin injector partly covered the surfaces). A typical FMR spectrum at 37 GHz and a frequency dependence of Py/LAO(5 u.c.)/STO sample are shown in Figs. 5a and 5b, respectively. The frequency dependence of $\Delta H$ exhibited a linear dependence with the frequency, and $\alpha$ was estimated to be $1.06\times10^{-3}$. Figure 5c shows the u.c. number dependence of $\alpha$. We observed the same $\alpha$ in all samples within experimental error, i.e. no damping enhancement. This indeed is consistent with the notion that the 2DEG is a very good spin conductor, or equivalently a very poor spin sink, in full support of the observation of a spin diffusion length of 300 nm. In a LAO/STO system, spin relaxation mechanism was studied via magnetoresistance measurements at low temperature[13], where it is reported that the spin relaxation mechanism at the LAO/STO interface was the D'yakonv-Perel type due to the Rashba spin-orbit interaction (SOI) induced by the large permittivity of the STO at low temperature ($\varepsilon_r > 20000$). Meanwhile, the Rashba SOI becomes weaker as decreasing the electric field perpendicular to the LAO/STO interface. In fact, the permittivity of the STO rapidly decreases with increasing temperature down to several hundreds (two orders of magnitude smaller) at RT[44,45]. We believe the realization of the long spin diffusion length in this study is attributed to the suppression of the Rashba field at RT.

In summary, we realized spin transport through a *d*-electron 2DEG at the LAO/STO interface at RT. Systematic control experiments supported the central claim. This achievement counters the conventional understanding of the physical properties of the *d*-electron-based 2DEGs, i.e., non-negligible spin orbit interaction and possible ferromagnetism can impede spin transport. Furthermore, this report pioneers a new approach to adding spintronics functions to the conductive oxide interface.

**Methods**

**Sample Fabrication:** The epitaxial LAO thin film on the STO(100) substrate was grown by pulsed laser deposition with a KrF excimer laser using a LAO single crystalline target. The commercially available STO substrates were $TiO_2$-terminated and by etching with buffered $NH_4F$-HF solutions (pH = 4.0) [M. Kawasaki, K. Takahashi, T. Maeda, R. Tsuchiya, M. Shinohara, O. Ishiyama, T. Yonezawa, M. Yoshimoto, and H. Koinuma, Science **266**, 1540 (1994).] and were then annealed at 1,050 ℃ for 1 hour in air. The temperatures and oxygen partial pressure were set to be 550 ℃ and $1\times10^{-3}$ Pa, respectively, during the deposition. Next, the samples were annealed in an oxygen pressure of 1 Pa at 900 ℃ for 30 minutes to eliminate oxygen vacancies. The film thickness was calibrated by x-ray reflectivity or by the intensity oscillation of reflection high energy electron diffraction. We have confirmed that the samples with 5 u.c. LAO thickness are conducting while those with 2 u.c. LAO thickness are insulating. The sheet resistance, sheet carrier mobility and Hall mobility of conducting samples agreed with the literature values. The samples for transport were equipped with a Py strip (27 nm in thickness) and a Pt wire (7.4 nm in thickness) or a Ta wire (7.9 nm in thickness) on top of the LAO(5 u.c.)/STO substrate by the electron beam lithography, electron beam evaporation and sputtering.

**Spin transport measurement:** Both sides of the edge of the Pt electrode were connected to a Nanovoltmeter with Cu wires, and an FMR of the Py was induced by the ESR system (Bruker, $TE_{102}$ cavity). The microwave frequency and power were set to 9.63 GHz and 200 mW, respectively, and a static magnetic field was applied to obtain the FMR (swept from 70 to 120 mT when the angle was set to be 0 and 180 degrees, and swept from 1350 to 1400 mT when the angle was set to be 90 degrees). All measurements were performed at room temperature.

**FMR measurement for the estimation of the Gilbert damping constant:** LAO(2, 5, 10)/STO substrates were prepared as described above, and Py (7 nm) was evaporated onto the substrate by electron beam deposition. The samples were set on a Ag/Cu coplanar waveguide located in an electromagnet. FMR measurements were performed from 5 to 40 GHz with the Ag/Cu coplanar waveguide and a magnetic coil at room temperature. The magnetic field was applied perpendicular to the Py surface to avoid unwanted damping due to a magnon-magnon scattering.

**Acknowledgement**

This research was supported in part by a Gran-in-Aid for Scientific Research from the Ministry of Education, Culture, Sports, Science and Technology (MEXT) of Japan, Innovative Area "Nano Spin Conversion Science" (No. 26103003), Scientific Research (S) "Semiconductor Spincurrentronics" (No. 16H0633), JSPS KAKENHI Grant (No. 16J00485), JSPS KAKENHI Grant (No. 25286506) and Elements Strategy Initiative to Form Core Research Center. One of the authors (R.O.) acknowledges JSPS Research Fellow.

**Figure captions**

**Figure 1 | Experimental set-up and current-voltage (*I-V*) characteristics of the Py-LAO/STO-detection electrodes. a,** Experimental concept and sample structure. Spin current is generated under the FMR and propagates at the LAO(5 u.c.)/STO interface from the Py to the NM (= Pt or Ta) wire. In the NM wire, the pure spin current is converted into a charge current, resulting in detection of an EMF. The *I-V* characteristics from the sample of **b,** Py-LAO/STO-Pt and **c,** Py-LAO/STO-Ta, respectively. The inset shows the measurement set-up.

**Figure 2 | Angular dependences of electromotive force of the inverse spin Hall effect. a-c,** The FMR spectrum and **d-f**, the EMFs of the Pt electrode sample (LAO, 5 u.c. and *L*, 890 nm) at 0, 90 and 180 degrees, respectively. The clear EMFs under the resonant point, $H_{FMR}$, can be seen. The amplitude of a non-negligible asymmetric component is equivalent to that of the ISHE-induced EMF, and the signals were deconvoluted by using a conventional fitting function. The origin of the asymmetric component is discussed in the main text. The symmetry of the EMFs is consistent with equation (1), thus suggesting that the EMFs are due to the ISHE of the Pt, i.e., the successful spin transport was achieved in the 2DEG.

**Figure 3 | Electromotive forces obtained from the Ta electrode sample. a-c,** The FMR spectrum of the Py and **d-f**, the EMFs of the Ta electrode sample (LAO, 5 u.c.) at 0, 90 and 180 degrees, respectively. Because the sign of the spin Hall angle of Ta is opposite from that of Pt, the EMFs of the ISHE are again consistent with equation (1), thereby providing further evidence for the successful spin transport in the 2DEG.

**Figure 4 | Magnitudes of the ISHE-induced electric current as a function of the gap length between the Py and the Pt or Ta.** The normalized ISHE-induced electric current $I_{\text{norm}} = V_{\text{ISHE}}/R|\theta_{\text{SHE}}^{\text{NM}}|\lambda_{\text{NM}}\tanh(d/2\lambda_{\text{NM}})$ is plotted by changing the gap length from the Py to the NM (= Pt or Ta), $L$. The dashed line is an exponential fitting line.

**Figure 5 | The Gilbert damping $\alpha$ of full-capping Py on the 2DEG at LAO/STO interfaces under spin pumping conditions. a,** An FMR spectrum of the Py on the LAO(5 u.c.)/STO at 37 GHz. **b,** Frequency dependence of the full width at half maximum of Py/LAO(5 u.c.)/STO showing a linear behavior consistent with the equation of the Gilbert damping. **c,** LAO thickness dependence of the Gilbert damping parameter, $\alpha$. No enhancement of $\alpha$ provides an evidence of efficient spin transport in the 2DEG. The error bars in the figure are the standard deviation.

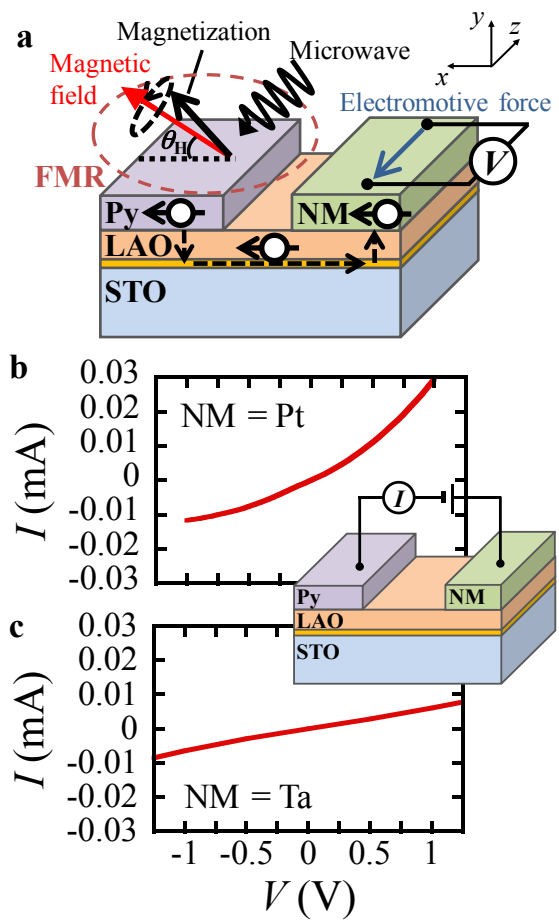

Fig. 1 R. Ohshima et al.

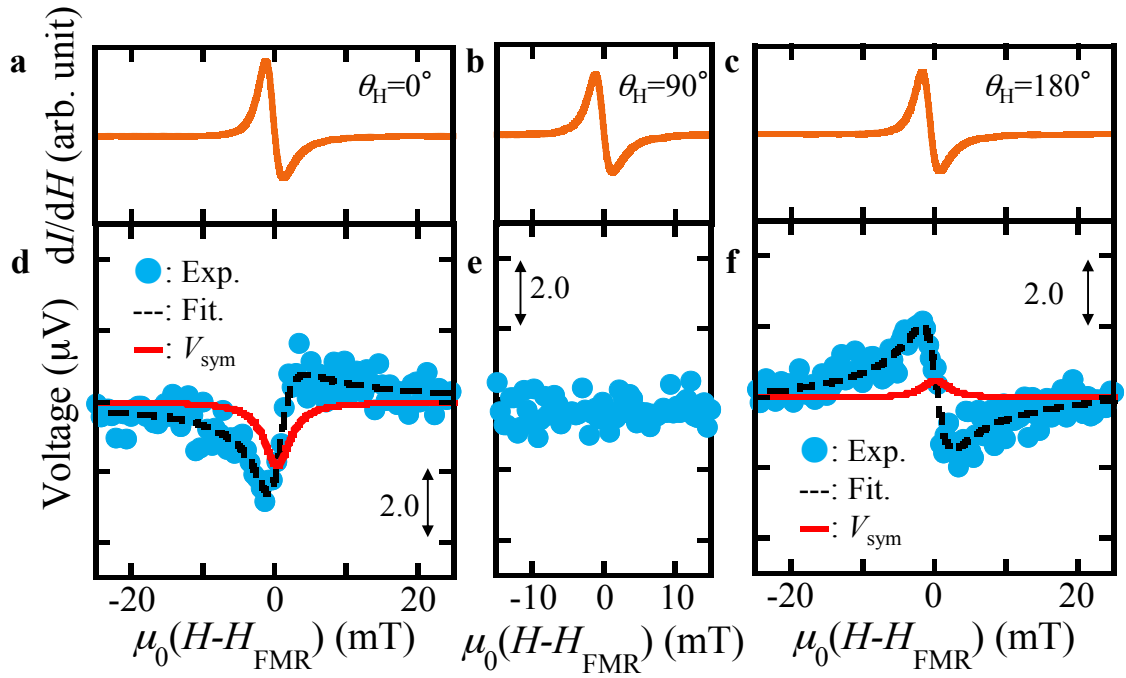

Fig. 2 R. Ohshima et al.

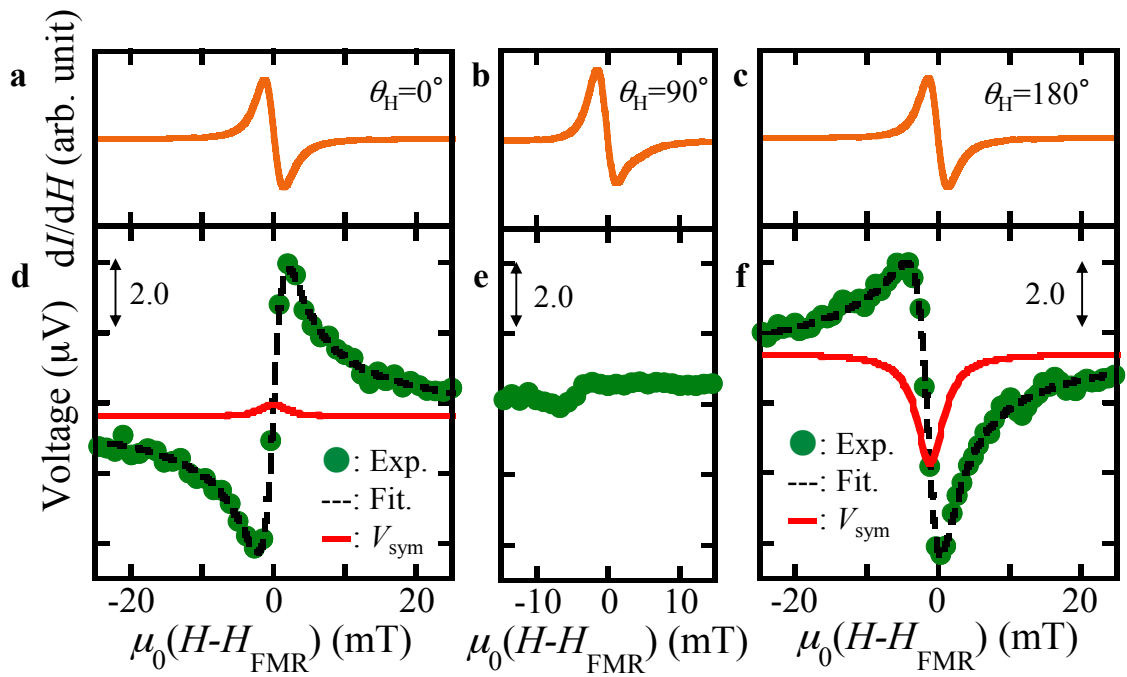

Fig. 3 R. Ohshima et al.

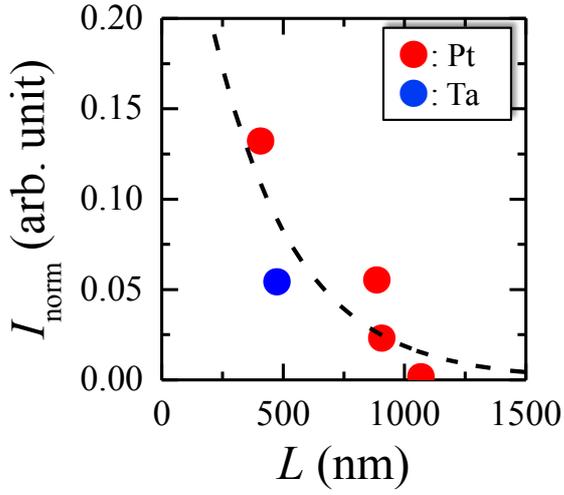

Fig. 4 R. Ohshima et.al.

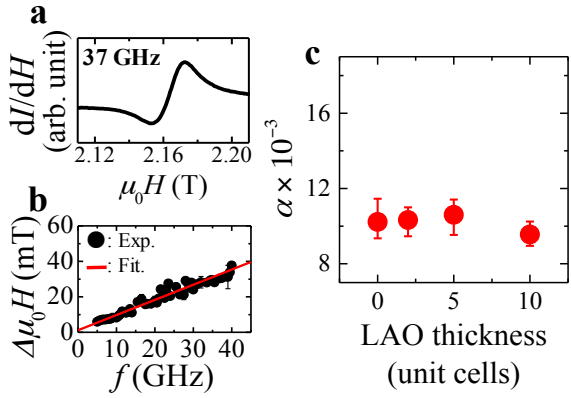

Fig. 5 R. Ohshima et al.

**Table and Table caption**

| Table I | Symmetric and asymmetric components of the electromotive forces obtained from the Pt and Ta detectors. | |
|---|---|---|
| **Electrode** | $V_{ISHE}$ | $V_{Asym}$ |
| Pt | -1.16 | -1.58 |
| Ta | -1.74 | -3.95 |

Parameters: inverse spin Hall effect voltage $V_{ISHE}$ (μV) and averaged asymmetric component $V_{Asym}$ (μV).

*Supplementary Information*

# Realization of *d*-electron spin transport at room temperature at a LaAlO$_3$/SrTiO$_3$ interface

Ryo Ohshima [1,2], Yuichiro Ando [1], Kosuke Matsuzaki [3], Tomofumi Susaki [3], Mathias Weiler [4,5], Stefan Klingler [4,5], Hans Huebl [4,5,6], Eiji Shikoh [7], Teruya Shinjo [1], Sebastian T. B. Goennenwein [4,5,6] and Masashi Shiraishi [1,*]

1. Department of Electronic Science and Engineering, Kyoto Univ., 615-8510 Kyoto, Japan.
2. Graduate School of Engineering Science, Osaka Univ., 560-8530 Toyonaka, Japan.
3. Secure Materials Center, Materials and Structures Lab., Tokyo Institute of Technology, 226-8503 Yokohama, Japan.
4. Walther-Meißner-Institut, Bayerische Akademie der Wissenschaften, 85748 Garching, Germany.
5. Physik-Department, Technische Universität München, 85748 Garching, Germany
6. Nanosystems Initiative Munich (NIM), Schellingstraße 4, 80799 München, Germany
7. Graduate School of Engineering, Osaka City Univ., 558-8585 Osaka, Japan.

**A. Fitting analysis of the electromotive force**

The electromotive force (EMF) under the ferromagnetic resonance (FMR) consists of symmetric and asymmetric components when a detector and a ferromagnet are electrically connected. A deconvolution of the spectrum was performed by using the equation given by

$$V(H) = V_{\text{sym}} \frac{\Gamma^2}{\mu_0^2(H - H_{\text{FMR}})^2 + \Gamma^2} + V_{\text{asym}} \frac{-2\Gamma\mu_0(H - H_{\text{FMR}})}{\mu_0^2(H - H_{\text{FMR}})^2 + \Gamma^2} + V_{\text{BG}}(H), \quad \text{(S1)}$$

where $V_{\text{sym}}$ and $V_{\text{asym}}$ are the voltages contribute to the EMF as symmetric and asymmetric component, respectively, $\Gamma$ is the full width at half maximum, $\mu_0 H_{\text{FMR}}$ is the resonance field of Py, and $V_{\text{BG}}$ is the linear back ground[1]. All EMFs in main text are fitted by this equation and the deconvolution into the symmetric and asymmetric parts shown in Fig. S1. The fitting parameters are listed in Table SI.

**B. Estimation of the spin current density injected into the LAO/STO interfaces in the spin transport sample**

The magnetization dynamics of the Py wire $\boldsymbol{M}(t)$ are described by the Landau-Lifshitz-Gilbert equation as

$$\frac{d\boldsymbol{M}(t)}{dt} = -\gamma\mu_0 \boldsymbol{H}_{\text{eff}} \times \boldsymbol{M}(t) + \frac{\alpha}{M_S} \boldsymbol{M}(t) \times \frac{d\boldsymbol{M}(t)}{dt}, \quad \text{(S2)}$$

where $\mu_0$, $\gamma$, $\boldsymbol{H}_{\text{eff}}$, $\alpha$ and $M_S$ are the vacuum permeability, the gyromagnetic ratio, the effective magnetic field, Gilbert damping constant and saturation magnetization of Py. When the FMR condition is met, the magnetization is resonantly excited out of equilibrium, thereby pumping a spin current from Py into the 2DEG in the LAO/STO. Given that the metal (Pt or Ta) detector strip is within the spin decay length in the 2DEG, the injected spins can reach the metal detector strip and induce an inverse spin Hall effect (ISHE) in the latter. The injected spin current density $j_S^{\text{Py}}$ can be expressed as

$$j_S^{Py} = \frac{\omega}{2\pi} \int_0^{\frac{2\pi}{\omega}} \frac{\hbar}{4\pi} g_r^{\uparrow\downarrow} \frac{1}{M_S^2} \left[ \mathbf{M}(t) \times \frac{d\mathbf{M}(t)}{dt} \right]_z dt$$

$$= \frac{g_r^{\uparrow\downarrow} \gamma^2 \mu_0^2 h_{MW}^2 \hbar \left[ \mu_0 M_S \gamma + \sqrt{(\mu_0 M_S)^2 \gamma^2 + 4\omega^2} \right]}{8\pi\alpha^2 [(\mu_0 M_S)^2 \gamma^2 + 4\omega^2]},$$

(S3)

where $\omega = 2\pi f = 6.05 \times 10^{10}$ rad/s is the cyclic frequency of the magnetization precession, $\hbar$ is the Dirac constant, $\mu_0 h_{MW} = 6.00 \times 10^{-2}$ mT is the amplitude of the microwave field under the microwave power of 200 mW and $g_r^{\uparrow\downarrow}$ is the real part of the mixing conductance[2,3]. $g_r^{\uparrow\downarrow}$ can be estimated from the difference of the FMR spectral width between Py/LAO/STO and Py/SiO$_2$ $(W_{Py/2DEG} - W_{Py/SiO_2})$:

$$g_r^{\uparrow\downarrow} = \frac{4\pi M_S d_{Py}^{NM}}{g\mu_B} (\alpha_{Py/2DEG} - \alpha_{Py/SiO_2}),$$

(S4)

where $g = 2$ is the g-factor, $\mu_B$ is the Bohr magneton and $d_{Py}^{Pt} = 27$ nm ($d_{Py}^{Ta} = 30$ nm) is the thickness of the Py wire in the case of NM (= Pt or Ta) electrode sample[3]. $\alpha_i = \sqrt{3}\gamma W_i / 2\omega$ ($i = $ Py/2DEG, Py/SiO$_2$) is the Gilbert damping constant of Py on LAO/STO and on SiO$_2$. $(W_{Py/2DEG} - W_{Py/SiO_2}) = 4.5 \times 10^{-1}$ mT was estimated from FMR spectrum in Fig. S2 and $\mu_0 M_S = 1.02$ T (0.98 T) was calculated from the resonance field $\mu_0 H_{FMR} = 95$ mT (98 mT) and Kittel's equation[4]. Consequently, we can estimate $g_r^{\uparrow\downarrow} = 1.78 \times 10^{-19}$ m$^{-2}$ (1.97×10$^{-19}$ m$^{-2}$) and $j_S^{Py} = 7.44 \times 10^{-10}$ Jm$^{-2}$ (8.43×10$^{-10}$ Jm$^{-2}$) from equations (S3) and (S4) in the cases of the Pt and Ta electrode samples, respectively. Note that the analysis was applied to the results of the spin transport devices, where the LAO/STO surface was partly covered by the Py.

## C. Another approach for estimating the spin diffusion length in the LAO/STO interface

For the simple case of a ferromagnet/metal (F/N) bilayer, the open-circuit ISHE voltage $V_{\text{ISHE}}$ induced by spin pumping can be calculated in an effective 1D model[5-7]:

$$V_{\text{ISHE}} = \frac{e\theta_{\text{SHE}}^{\text{N}} \lambda_{\text{N}} \tanh(d_{\text{N}}/2\lambda_{\text{N}})}{d_{\text{N}}\sigma_{\text{N}} + d_{\text{F}}\sigma_{\text{F}}} l_{\text{N}} \eta P \frac{\omega}{2\pi} g_r^{\uparrow\downarrow} \sin^2\Theta_{\text{res}}, \qquad (S5)$$

where $\eta = \left[1 + 2g_r^{\uparrow\downarrow} \rho_{\text{N}} \lambda_{\text{N}} \frac{e^2}{h} \coth\left(\frac{d_{\text{N}}}{\lambda_{\text{N}}}\right)\right]^{-1}$ is a correction factor taking into account spin current backflow[8], $P \sim 1$ is introduced to correct for the ellipticity of the magnetization precession, and $\Theta_{\text{res}}$ is the magnetization precession cone angle in resonance. The term $P \frac{\omega}{2\pi} g_r^{\uparrow\downarrow} \sin^2\Theta_{\text{res}}$ in (S4) is the spin current pumped across the F/N interface in the FMR[6].

For our experiment, the spin pumping formalism enshrined in Eq. (S5) can only be used as a first approximation. Indeed, for F/N1/N2 trilayers, Boone et al.[9,10] have shown that a more evolved treatment is mandatory to capture the spin current transfer across two consecutive interfaces and spin dephasing in the different layers. Our lateral F/2DEG/N nanostructures feature 2 different interfaces (F/2DEG and 2DEG/N), with highly anisotropic spin transport properties in the 2DEG. We here nevertheless use (S5) to estimate an effective spin transfer efficiency which we call effective mixing conductance $g_{\text{F/2DEG/N}}^{\uparrow\downarrow}$. The latter can be interpreted as measure for the spin transfer efficiency from F to N via the 2DEG. In other words, the spin transfer across the F/2DEG and the 2DEG/N interfaces, as well as spin diffusion and dephasing on the way, all are lumped together in $g_{\text{F/2DEG/N}}^{\uparrow\downarrow}$. We furthermore assume that electrical transport in N is dominant, and use

$$V_{\text{ISHE}} = \frac{e\theta_{\text{SHE}}^{\text{N}} \lambda_{\text{N}} \tanh(d_{\text{N}}/2\lambda_{\text{N}})}{d_{\text{N}}\sigma_{\text{N}}} l_{\text{N}} \eta \frac{\omega}{2\pi} g_{\text{F/2DEG/N}}^{\uparrow\downarrow} \sin^2\Theta_{\text{res}}, \qquad (S6)$$

to calculate $g_{\text{F/2DEG/N}}^{\uparrow\downarrow} \approx 1.9\times10^{17}$ m$^{-2}$ from $V_{\text{ISHE}}^{\text{Pt}} = 1.16$ μV, $\theta_{\text{SHE}}^{\text{Pt}} = 0.1$, $\lambda_{\text{Pt}} = 7.3$ nm, $l_{\text{Pt}} = 900$ μm, $d_{\text{Pt}} = 7.4$ nm, $\sigma_{\text{Pt}} = 2.1\times10^6$ 1/Ωm, $\omega = 6.05\times10^{10}$ s$^{-1}$, $\eta = 0.22$ and $\Theta_{\text{res}} = 2h_{\text{MW}}/\Delta H_{FWHM}$ inferred from the RF magnetic field $\mu_0 h_{\text{MW}} = 6.00\times10^{-2}$ mT at a microwave power of 200 mW and the full width at half maximum $\mu_0 \Delta H_{FWHM} = 4.13$ mT. The $g^{\uparrow\downarrow}_{F/2DEG/N}$ value thus obtained is about two orders of magnitude smaller than the typical $g^{\uparrow\downarrow}_r$ values reported $\approx 10^{19}$ m$^{-2}$ for plain F/N interfaces[11]. This makes sense, since $g^{\uparrow\downarrow}_{F/2DEG/N}$ contains the spin transfer across two consecutive interfaces together with spin diffusion and dephasing in the 2DEG, as already mentioned. Thus, $g^{\uparrow\downarrow}_{F/2DEG/N} \ll g^{\uparrow\downarrow}_r$ is expected. The spin diffusion length of LAO/STO interfaces $\lambda_{2DEG} \approx 360$ nm can be estimated from the relationship between spin current from Py to 2DEG $j^{Py}_s$ and from 2DEG to Pt $j^{Pt}_s$: $j^{Pt}_s = j^{Py}_s \exp(-L/\lambda_{2DEG})$, where $j^{Py}_s \approx 1\times 10^{-9}$ Jm$^{-2}$ and $j^{Pt}_s = 8.24 \times 10^{-11}$ Jm$^{-2}$ are estimated from the relationship $j_s = \frac{\hbar}{2} P \frac{\omega}{2\pi} g^{\uparrow\downarrow}_r \sin^2 \Theta_{res}$. Estimated value is very close to that in the main text.

**D. Diffusion constant of the LAO/STO interfaces**

The diffusion constant was estimated by using $D = v_F^2 \tau/2$, $v_F = \sqrt{2E_F/m^*} = \hbar\sqrt{2\pi n}/m^*$ and $\tau = m^*\sigma/e^2 n$, where $E_F$ is the Fermi level, $m^*$ is the effective mass, $T$ is the temperature, $\sigma = 2.58\times 10^{-5}$ S is the sheet conductance, $e$ is the charge and $n = 3.90\times 10^{13}$ cm$^{-2}$ is a sheet carrier concentration[12]. $\sigma$ and $n$ were estimated by the van der Pauw method, and $m^* = 2.1 m_e$ was obtained from the referenced values[13].

**Figure captions**

**Figure S1 | Deconvolution of the electromotive forces obtained from the Pt and Ta**

**detectors.** Experimentally observed electromotive forces are divided into symmetric ($V_{sym}$) and asymmetric components ($V_{asym}$) by equation (S1).

**Figure S2 | Comparison of the FMR spectral width of Py between the cases of Py on 2DEG and Py on SiO$_2$.** The FMR spectrum of Py/LAO/STO and Py/SiO$_2$, respectively.

**Figures**

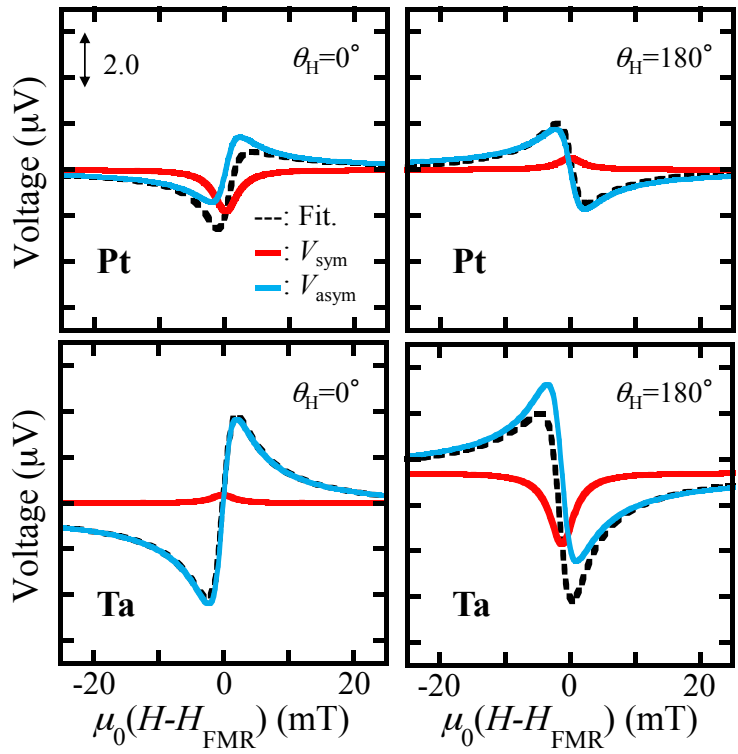

Fig. S1 R. Ohshima et al.

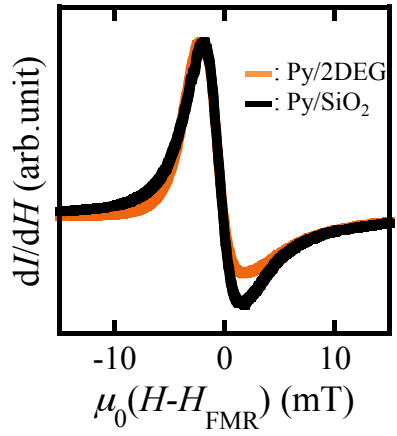

Fig. S2 R. Ohshima et al.

**Table and Table Caption**

| Table SI | Fitting parameters of electromotive forces | | | | |
|---|---|---|---|---|---|
| **Electrode** | $\theta_H$ | $V_{sym}$ | $V_{asym}$ | $\Gamma$ | $\mu_0 H_{FMR}$ |
| Pt | 0 | -1.82 | -1.42 | 2.19 | 95.2 |
| Pt | 180 | 0.50 | 1.74 | 2.25 | 95.0 |
| Ta | 0 | 0.35 | -4.04 | 2.17 | 97.6 |
| Ta | 180 | -3.12 | 3.86 | 2.25 | 97.9 |

Parameters: symmetric and asymmetric component of EMF $V_{sym}$ (μV) and $V_{asym}$ (μV), the full width at half maximum $\Gamma$ (mT) and resonance field $\mu_0 H_{FMR}$ (mT)